\documentclass[aps,twocolumn,prd,showpacs,floatfix]{revtex4}
\usepackage{bm}
\usepackage{amsmath}
\usepackage{nicefrac}
\usepackage{amsthm}
\usepackage{amssymb}
\usepackage{graphicx}
\usepackage{color}

\newcommand{\eps }{\varepsilon}

\newcommand{\appropto}{\mathrel{\vcenter{
  \offinterlineskip\halign{\hfil$##$\cr
    \propto\cr\noalign{\kern2pt}\sim\cr\noalign{\kern-2pt}}}}}

\begin{document}

\title{Axion-induced effects in atoms, molecules and nuclei: parity non-conservation, anapole moments, electric dipole moments, and spin-gravity and spin-axion momentum couplings}

\date{\today}
\author{Y.~V.~Stadnik}
\affiliation{School of Physics, University of New South Wales, Sydney 2052,
Australia}
\author{V.~V.~Flambaum}
\affiliation{School of Physics, University of New South Wales, Sydney 2052,
Australia}

\begin{abstract}
We show that the interaction of an axion field, or in general a pseudoscalar field, with the axial-vector current generated by an electron through a derivative-type coupling can give rise to a time-dependent mixing of opposite-parity states in atomic and molecular systems. Likewise, the analogous interaction of an axion field with the axial-vector current generated by a nucleon can give rise to time-dependent mixing of opposite-parity states in nuclear systems. This mixing can induce oscillating electric dipole moments, oscillating parity non-conservation effects and oscillating anapole moments in such systems. By adjusting the energy separation between the opposite-parity states of interest to match the axion mass energy, axion-induced experimental observables can be enhanced by many orders of magnitude. Oscillating atomic electric dipole moments can also be generated by axions through hadronic mechanisms, namely the P,T-violating nucleon-nucleon interaction and through the axion-induced electric dipole moments of valence nucleons, which comprise the nuclei. The axion field is modified by the Earth's gravitational field. The interaction of the spin of either an electron or nucleon with this modified axion field leads to axion-induced observable effects. These effects, which are of the form $\mathbf{g} \cdot \mathbf{\sigma}$, differ from the axion-wind effect, which has the form $\mathbf{p}_{\textrm{a}} \cdot \mathbf{\sigma}$.

\end{abstract}


\maketitle 

\section{Introduction}
\label{Sec:Intro}

The strong CP problem, which seeks to explain why quantum chromodynamics (QCD) does not appear to violate the combined charge-parity (CP) symmetry, remains one of the most important outstanding problems in fundamental physics to date. One possible resolution of the strong CP problem is that the QCD CP-symmetry breaking parameter $\theta$ becomes unobservable if at least one of the quarks is massless (see, for instance, review \cite{Kim_RMP_10}). However, there does not appear to be empirical evidence to date that any of the quarks in the Standard Model (SM) are massless and so this resolution mechanism seems unlikely. An alternative explanation of the strong CP problem is offered by the Peccei-Quinn (PQ) theory, in which an additional global U($1$) symmetry, known as the PQ symmetry, is introduced into the SM QCD Lagrangian and is subsequently broken both spontaneously and explicitly \cite{PQ77a,PQ77b}. See also Refs.~\cite{Weinberg78x,Wilczek78x,Kim79x,Zakharov80x,Zhitnitsky80x,Srednicki81x}. The breaking of the PQ symmetry gives rise to a pseudoscalar pseudo-Nambu-Goldstone boson, known as the axion, being born from the QCD vacuum and causes the $\theta$ parameter to become effectively zero, thus in principle alleviating the strong CP problem. 

Another outstanding problem of great importance in contemporary physics is that of dark matter, specifically cold dark matter (CDM), the existence of which is generally accepted on the basis of overwhelming astrophysical evidence (see e.g.~Refs.~\cite{Zwicky37,Rubin70,Rubin80,deBlok01,Koopmans03,Ostriker03,Dekel05,Minchin05,Clowe06,Vikhlinin06,Weinberg06,Massey07,Governato10,Jorg12}), but the composition of which is much less clear. We do know, however, that the matter-energy content of the universe is overwhelming dominated by CDM ($\sim 23 \%$) and dark energy ($\sim 73 \%$), with only a few percent attributable to baryonic matter (see e.g.~Ref.~\cite{Spergel07}). 
There are several possible candidates for CDM, including weakly interacting massive particles (WIMPs), super-weakly interacting massive particles (super-WIMPs), massive astrophysical compact halo objects (MACHOs), such as primordial black holes, and axions (see, for instance, Ref.~\cite{PDG12} and the plethora of references therein for further details of the properties of and searches for these particles). 

In the present work, we restrict our attention to axions. For more details to the ensuing discussion of axion theories for CDM, we refer the reader to Refs.~\cite{PQ77a,PQ77b,PDG12,Wilczek83,Abbott83,Dine83,Turner86}. Axion theories for CDM predict that significant quantities of axionic matter may have been formed shortly after the Big Bang. At a sufficiently large temperature (well above the QCD critical temperature), the axion is massless and the axion field can have essentially any value, parametrised by the misalignment angle $\theta_{\textrm{i}}$. As the axion plasma cooled to below the QCD critical temperature, the axion attained a mass and, since the axion field was initially unlikely to be near the minimum of the potential for the field, the axions dissipated most of their kinetic energy as they fell into the nearest potential minima via the so-called misalignment mechanism. If axions have sufficiently low mass that no other decay modes were possible during the misalignment mechanism, then, due to the bosonic nature of axions, the universe would have been pervaded by a Bose-Einstein condensate (BEC) of primordial axions possessing very little kinetic energy. The suppression of other decay modes means that a remnant axionic background field should still exist at present and, at least in principle, should be detectable.  

While there exist numerous schemes for the detection of WIMPs, for instance, there are comparatively few detection schemes for axionic dark matter. One of the main detection schemes for axions involves detecting the conversion of axions into photons in a microwave cavity, which is permeated by a homogeneous magnetic field (see e.g.~Refs.~\cite{Sikivie85,CAST07,ADMX10}). Another popular detection scheme for axions involves measuring the axio-electric effect, which is the ionisation of (usually atomic) matter by axions (as opposed to by photons in the photo-electric effect) (see e.g.~Refs.~\cite{Avignone87,Pospelov08,Avignone09a,Avignone09b,Derevianko10,Dzuba10,Derbin12,Derbin13}). There also exist variants of the axio-electric effect involving Primakov conversion (see e.g.~Ref.~\cite{CDMS09}), as well as Compton and bremsstrahlung-like processes involving axions (see e.g.~Ref.~\cite{Derbin11}). Bounds obtained from astrophysical data assist us in axion CDM searches by ruling out a large region of the allowed values of axion parameters \cite{Dicus78,Raffelt08}.

More recently, Ref.~\cite{Graham11} suggested to search for axionic CDM through energy shifts in atomic systems arising from the coupling of axions to gluons, a process which can also give rise to a non-zero nucleon electric dipole moment (EDM). In Refs.~\cite{Victor13,Graham13}, another method is suggested to search for axionic CDM using atomic systems. The essence of this method is as follows. The Solar System rotates about the centre of the Milky Way Galaxy with a circular speed of $v_{\textrm{rot}} \approx 240$ km/s \cite{PDG12}. Thus our Solar System should be passing through an axion `wind'.
Note that the present-day background axionic field must invariably differ from the primordial axionic BEC, which formed shortly after the Big Bang, due to gravitational perturbations.
 As an estimate of the velocities of the axions comprising the background axionic field due to thermal motion, it is usual to assume that the root-mean-squared velocity of the axions is given by $v_{\textrm{rms}} = \sqrt{3/2}~v_{\textrm{rot}} \sim 290$ km/s \cite{Catena12}. 
 The background axion field can interact with the axial-vector current generated by electrons and nucleons through a derivative-type coupling (see e.g.~Refs.~\cite{Pospelov08,Derevianko10,Dzuba10}). The time-dependent potential arising from the spatial components of this interaction is proportional to $\mathbf{p}_{\mathrm{a}} \cdot \mathbf{\sigma}_{\mathrm{\lambda}}$, where $\mathbf{p}_{\mathrm{a}}$ is the momentum of an axion comprising the axionic background field relative to an observer on Earth and $\mathbf{\sigma}_{\mathrm{\lambda}}$ is the spin operator for an electron ($\lambda=e$) or nucleon ($\lambda=N$) in the atomic, molecular or nuclear system of interest. Thus this time-dependent interaction is of the same form as that due to a time-dependent magnetic field applied to an atomic, molecular or nuclear system, and can give rise to energy level shifts in the species under consideration. These energy shifts can be measured in principle. Such a method can probe previously inaccessible regions of the axion parameter space \cite{Graham11,Graham13}. 

In this paper, we show that this same interaction of an axion field, or in general a pseudoscalar field, with the axial-vector current generated by an electron field through a derivative-type coupling can also give rise to a time-dependent mixing of opposite-parity states in atomic and molecular systems. An analogous effect also arises in nuclear systems due to the interaction of an axion field with the axial-vector current generated by a nucleon field through a derivative-type coupling of the same form. This mixing can induce oscillating EDMs, oscillating parity non-conservation (PNC) effects and oscillating anapole moments in such systems, the first of which can be measured by the methods discussed in Refs.~\cite{Graham11,Graham13}. We suggest that the first two of these effects can be measured through the application of a static electric field to the system of interest, and derive expressions for such axion-induced EDMs in group I elements and systems possessing a single nearly degenerate pair of opposite-parity states. By adjusting the energy separation between the opposite-parity states of interest to match the axion mass energy, axion-induced experimental observables can be enhanced by many orders of magnitude. This is essentially a resonance phenomenon. Measurements of these effects permit either the determination of or the placing of limits on important physical axion parameters. 
We consider oscillating atomic EDMs that can be generated by axions through hadronic mechanisms, namely the P,T-violating nucleon-nucleon interaction and through the axion-induced EDMs of valence nucleons, the latter of which was considered in Refs.~\cite{Graham11,Graham13}, and derive corresponding expressions for the axion-induced EDM for $^{199}$Hg, which at present provides the most sensitive probe for static EDM measurements in diamagnetic atoms \cite{Romalis09,Romalis13A}, and $^{225}$Ra (also $^{223}$Rn and $^{223}$Ra) which can offer a several order-of-magnitude enhancement in EDM magnitude over that for $^{199}$Hg. We also show that the interaction of the spin of either an electron or nucleon, which also interacts with an axion field, with the gravitational field gradient of a gravitating body can give rise to axion-induced observable effects. These effects, which are of the form $\mathbf{g} \cdot \mathbf{\sigma}_{\mathrm{\lambda}}$, differ from the axion-wind effect, which has the form $\mathbf{p}_{\mathrm{a}} \cdot \mathbf{\sigma}_{\mathrm{\lambda}}$. 

The structure of this paper is as follows. In Sec.~\ref{Sec:Theory}, we present and derive necessary theory, showing how the mixing of opposite-parity states can arise in atomic and molecular systems due to the interaction of electrons with a background axionic field, and likewise in nuclear systems through the interaction of nucleons with a background axionic field. We show how this mixing can induce oscillating EDMs, oscillating PNC effects and oscillating anapole moments in such systems, and derive corresponding expressions for such axion-induced EDMs in Group I elements and systems with a single nearly degenerate pair of opposite-parity states. In Sec.~\ref{Sec:Overview}, we briefly recapitulate the essence of one particular Stark-interference technique variant in atomic and molecular experiments designed to measure the static mixing of opposite-parity states induced by the neutral weak interaction. In Sec.~\ref{Sec:pnc}, we show how the application of a static electric field can be used to measure the oscillating EDMs and PNC effects of Sec.~\ref{Sec:Theory}. Then in Sec.~\ref{Sec:hadronic}, we consider oscillating atomic EDMs that can be generated by axions through hadronic mechanisms and derive corresponding expressions for the axion-induced EDMs of $^{199}$Hg and $^{225}$Ra. In Sec.~\ref{Sec:grav}, we show that the interaction of the spin of either an electron or nucleon with an axion field, modified by the gravitational field of a massive body, can give rise to axion-induced observable effects. These effects, which are of the form $\mathbf{g} \cdot \mathbf{\sigma}$, differ from the axion-wind effect, which has the form $\mathbf{p}_{\textrm{a}} \cdot \mathbf{\sigma}$. Finally, Sec.~\ref{Sec:Concl} presents our conclusions.

Note that, unless explicitly stated, we employ the natural units $\hbar=c=1$ hereafter. We also employ the metric signature $(+---)$ for flat, Minkowskian spacetime in this work, as well as the Einstein summation convention over repeated indices, which run over $\mu=0,1,2,3$.

\section{Theory}
\label{Sec:Theory}

The axion is a pseudoscalar particle and so must satisfy the Klein-Gordon equation, which in flat spacetime reads \cite{LL4}
\begin{equation}
\label{KG_eqn}
\left( \partial_{\mu} \partial^{\mu} + m^2 \right) \phi\left(\mathbf{r},t\right) = 0 ,
\end{equation}
where $\phi\left(\mathbf{r},t\right)$ is the axion field, which we assume to be classical and hence real. The solution to Eq.~(\ref{KG_eqn}) thus reads
\begin{equation}
\label{axion_field}
\phi\left(\mathbf{r},t\right) = a_{0} \cos\left(\mathbf{p}_{\mathrm{a}} \cdot \mathbf{r} - \varepsilon_{\mathrm{a}} t + \eta \right) ,
\end{equation}
where $\mathbf{p}_{\mathrm{a}}$ is the momentum of an axion, which comprises the background axionic field, relative to an observer on Earth, $\eta$ is a phase factor that depends on the initial conditions and $\varepsilon_{\mathrm{a}}$ is the energy of an axion particle, which is given by the following dispersion relation ($m_{\mathrm{a}}$ is the axion mass):
\begin{equation}
\label{ax_disp_reln}
\varepsilon_{\mathrm{a}} = \sqrt{\left|\mathbf{p}_{\mathrm{a}}\right|^{2} + m_{\mathrm{a}}^{2}} .
\end{equation}
The stress-energy tensor for the axion field is given by \cite{Goldstein02}
\begin{equation}
\label{SE_tensor}
T_{\mu}^{\nu} = \frac{\partial \mathcal{L}_{\mathrm{KG}}}{\partial \left(\partial_{\nu} \phi \right)} \left(\partial_{\mu} \phi \right) - \mathcal{L}_{\mathrm{KG}} \delta_{\mu}^{\nu} , 
\end{equation}
where $\mathcal{L}_{\mathrm{KG}}$ is the Klein-Gordon Lagrangian density given by \cite{Griffiths08}
\begin{equation}
\label{L_KG}
\mathcal{L}_{\mathrm{KG}} = \frac{1}{2}\left[\left(\partial_{\mu} \phi \right) \left(\partial^{\mu} \phi \right) - m^2 \phi^{2} \right] .
\end{equation}
From Eqs.~(\ref{SE_tensor}) and (\ref{L_KG}), we find
\begin{equation}
\label{SE_tensor2}
T_{\mu \nu} =  \left(\partial_{\mu} \phi \right) \left(\partial_{\nu} \phi \right) - \frac{1}{2} \left(\partial_{\rho} \phi \right) \left(\partial^{\rho} \phi \right) g_{\mu \nu}  + \frac{m^2 \phi^{2}}{2} g_{\mu \nu} . 
\end{equation}
Substituting Eqs.~(\ref{axion_field}) and (\ref{ax_disp_reln}) into Eq.~(\ref{SE_tensor2}), we find the energy density associated with the axion field to be
\begin{align}
\label{energy_density}
T_{00} &= \frac{a_{0}^{2}}{2} \left[ m_{\mathrm{a}}^{2} + 2 \left|\mathbf{p}_{\mathrm{a}}\right|^{2} \sin^{2}\left(\mathbf{p}_{\mathrm{a}} \cdot \mathbf{r} - \varepsilon_{\mathrm{a}} t + \eta \right) \right]  \notag \\
&= \frac{a_{0}^{2}}{2} \left( m_{\mathrm{a}}^{2} + \left|\mathbf{p}_{\mathrm{a}}\right|^{2} \right) ,
\end{align}
where in the second line of (\ref{energy_density}) we have have taken the time average.

The background axionic field can interact with the axial-vector current generated by an electron or nucleon (or any SM fermion in general), with the corresponding Lagrangian interaction density given by (see e.g.~Refs.~\cite{Pospelov08,Derevianko10,Dzuba10})
\begin{equation}
\label{L_int}
\mathcal{L}_{\mathrm{int}} = - \frac{\partial_{\mu} \phi}{f_{\textrm{a}}} \bar{\psi}\gamma^{\mu}\gamma^{5}\psi ,
\end{equation}
where $\psi$ is either the Dirac electron or nucleon field, $\bar{\psi} \equiv \psi^{\dagger} \gamma^{0}$ is the corresponding Dirac adjoint field and $f_{\textrm{a}}$ is the reciprocal of the coupling constant for the given interaction. Since the speed of the background axion field relative to an observer on the Earth, the typical speed of an electron in an atom or molecule, and the typical speed of a nucleon in a nucleus are all $\ll 1$, the interaction of interest is a non-relativistic one and so we can take the non-relativistic limit of Eq.~(\ref{L_int}). The temporal component of (\ref{L_int}) gives rise to the following partial interaction Hamiltonian in the non-relativistic limit (see e.g.~Ref.~\cite{Pospelov08}):
\begin{align}
\label{H_int_NRL_temp}
H_{\textrm{int}}^{\textrm{temp}} \left(t\right) &= \frac{\partial_{t} \phi}{f_{\textrm{a}}} \frac{\mathbf{p}_{\mathrm{\lambda}} \cdot \mathbf{\sigma}_{\mathrm{\lambda}}}{m_{\mathrm{\lambda}}} \notag \\
&= \frac{a_{0} \varepsilon_{\mathrm{a}} }{f_{\textrm{a}}} \frac{\mathbf{p}_{\mathrm{\lambda}} \cdot \mathbf{\sigma}_{\mathrm{\lambda}}}{m_{\mathrm{\lambda}}} \sin\left(\mathbf{p}_{\mathrm{a}} \cdot \mathbf{r} - \varepsilon_{\mathrm{a}} t + \eta \right) ,
\end{align}
where $\mathbf{p}_{\mathrm{\lambda}}$ is the momentum operator for an electron ($\lambda=e$) or nucleon ($\lambda=N$) in the atomic, molecular or nuclear system of interest, $\mathbf{\sigma}_{\mathrm{\lambda}}$ is the spin operator for the fermion of interest, $m_{\mathrm{\lambda}}$ is the fermion mass, and we have used Eq.~(\ref{axion_field}) in the second line of (\ref{H_int_NRL_temp}). The spatial components of (\ref{L_int}) give rise to the following partial interaction Hamiltonian in the non-relativistic limit:
\begin{align}
\label{H_int_NRL_spat}
H_{\textrm{int}}^{\textrm{spat}} \left(t\right) &= \frac{\left(\mathbf{\nabla} \phi \right) \cdot \mathbf{\sigma}_{\mathrm{\lambda}}}{f_{\textrm{a}}} \notag \\
&= - \frac{a_{0} \sin\left(\mathbf{p}_{\mathrm{a}} \cdot \mathbf{r} - \varepsilon_{\mathrm{a}} t + \eta \right)}{f_{\textrm{a}}} \mathbf{p}_{\mathrm{a}} \cdot \mathbf{\sigma}_{\mathrm{\lambda}} ,
\end{align}
where $\mathbf{p}_{\mathrm{a}}$ is the momentum of an axion comprising the background axionic field relative to an observer on Earth, and we have used Eq.~(\ref{axion_field}) in the second line of (\ref{H_int_NRL_spat}). Thus the time-dependent interaction (\ref{H_int_NRL_spat}) is of the same form as that due to a time-dependent magnetic field applied to an atomic, molecular or nuclear system, and can give rise to energy level shifts in the species under consideration. This is the axion-wind effect, which was considered in Refs.~\cite{Victor13,Graham13}. Note that the effective magnetic field for the case of nucleons is given by $B_{N}^{\textrm{eff}} (t) = \frac{H_{\textrm{int}} (t)}{\mu_{N}}$, where $\mu_N = \frac{e}{2m_{p}}$ is the nuclear magneton, while the effective magnetic field for the case of electrons is given by $B_{e}^{\textrm{eff}} (t) = \frac{H_{\textrm{int}} (t)}{\mu_B}$, where $\mu_B= \frac{e}{2m_{e}}$ is the Bohr magneton. Since $\mu_N \ll \mu_B$, $B_{N}^{\textrm{eff}} (t) \gg B_{e}^{\textrm{eff}} (t)$, which implies that a larger signal-to-noise ratio should be achievable for the case of nucleons. We shall return to the axion-wind effect in Sec.~\ref{Sec:grav}.

Note that the interaction described by Eq.~(\ref{H_int_NRL_temp}) is a P-odd interaction, just like the the neutral weak interaction between an atomic nucleus and an orbiting electron in the non-relativistic limit \cite{Bouchiat74}. Consequently, the interaction (\ref{H_int_NRL_temp}) can give rise to time-dependent mixing of opposite-parity states in atoms, molecules and nuclei. In order to see this, we first consider for simplicity a two-level subspace of either an atomic or molecular system spanned by two arbitrary, opposite-parity eigenstates $\left| A \right>$ and $\left| B \right>$, before proceeding to full calculations. For the resonance phenomenon described at the end of the section, the two-level approximation is likely to be a very good approximation in most cases. The atomic or molecular system is at rest in the Earth's frame of reference (that is, $\mathbf{r}$ is constant). Also, since $v_{\textrm{a}} \ll 1$, the dispersion relation (\ref{ax_disp_reln}) gives $\varepsilon_{\textrm{a}} \approx m_{\textrm{a}}$. So we can write (\ref{H_int_NRL_temp}) as follows
\begin{equation}
\label{H_int_NRL_temp1}
H_{\textrm{int}}^{\textrm{temp}} \left(t\right) = \frac{a_{0} m_{\mathrm{a}}}{f_{\textrm{a}}} \frac{\mathbf{p}_{\mathrm{e}} \cdot \mathbf{\sigma}_{\mathrm{e}}}{m_{\textrm{e}}} \cos\left(m_{\mathrm{a}} t + \eta' \right) ,
\end{equation}
where we have redefined the phase factor to be $\eta'$. Note that (\ref{H_int_NRL_temp1}) is a pseudoscalar interaction and so can only mix states with the same values of $j$ and $j_{\textrm{z}}$. Using the operator identity $\mathbf{p}_{\mathrm{e}} = i m_{\textrm{e}} \left[H,\mathbf{r}_{\mathrm{e}}\right]$, where $H$ is the non-relativistic atomic or molecular Hamiltonian, we find
\begin{align}
\label{M_elt_1}
&\left< A \right| H_{\textrm{int}}^{\textrm{temp}} \left(t\right) \left| B \right>  = \frac{i a_{0} m_{\mathrm{a}}}{f_{\textrm{a}}} \cos\left(m_{\mathrm{a}} t + \eta ' \right)  \notag \\ 
&\times \left[ \left(\varepsilon_A - \varepsilon_B \right) \left< A \right| \mathbf{r}_{\mathrm{e}} \cdot \mathbf{\sigma}_{\mathrm{e}} \left| B \right> - \left< A \right| \mathbf{r}_{\mathrm{e}} \cdot \left[H,\mathbf{\sigma}_{\mathrm{e}}\right] \left| B \right> \right] ,
\end{align}
where $\mathbf{r}_{\mathrm{e}}$ is the electron position operator. In the non-relativistic limit, the commutator $\left[H,\mathbf{\sigma}_{\mathrm{e}}\right]$ vanishes in the absence of external interactions. Hence we find that
\begin{equation}
\label{M_elt_5}
\left< A \right| H_{\textrm{int}}^{\textrm{temp}} \left(t\right) \left| B \right>  = i H_{A} \cos\left(m_{\mathrm{a}} t + \eta ' \right) ,
\end{equation}
where
\begin{equation}
\label{H_A}
H_{A} = \frac{a_{0} m_{\mathrm{a}} \left(\varepsilon_A - \varepsilon_B \right) \left< A \right| \mathbf{r}_{\mathrm{e}} \cdot \mathbf{\sigma}_{\mathrm{e}} \left| B \right>}{ f_{\textrm{a}} } .
\end{equation}
Note that the matrix element (\ref{M_elt_5}) between the opposite-parity eigenstates $\left| A \right>$ and $\left| B \right>$ scales linearly with the energy difference between these states and so for a nearly degenerate pair of opposite-parity states is very small. In such cases, relativistic calculations are needed to find the matrix elements, since the spin-dependent relativistic corrections cannot be neglected and may even give the dominant contribution to the matrix element (\ref{M_elt_5}) (see e.g.~the commutator term in Eq.~(\ref{M_elt_1}) which does not contain the small energy difference). Full relativistic many-body numerical calculations require sophisticated computer codes and will be performed in a separate publication. In this paper, we perform analytical estimates only.

In the presence of the off-diagonal interaction (\ref{M_elt_5}), the Hamiltonian for the two-level subspace spanned by the $\left| A \right>$ and $\left| B \right>$ parity eigenstates thus reads
\begin{equation}
\label{Ham_new}
H \left(t \right) =
\left[ \begin{array}{cc}
\varepsilon_{A} & iH_{A} \cos\left(m_{\mathrm{a}} t  \right) \\
-iH_{A} \cos\left(m_{\mathrm{a}} t  \right) & \varepsilon_{B} 
\end{array} \right] , 
\end{equation}
where we have set the phase factor $\eta ' = 0$ from now on without loss of generality, unless explicitly written otherwise. 
In the interaction picture \cite{Sakurai}, the unperturbed system wavefunction projection onto the two-level subspace of interest reads
\begin{equation}
\label{WF}
\left| \psi (t)\right> = c_{A}(t) e^{-i \varepsilon_{A}t} \left| A \right> + c_{B}(t) e^{-i \varepsilon_{B}t} \left| B \right> ,
\end{equation}
from which follow the following coupled differential equations:
\begin{equation}
\label{time_evoln_int_Cn}
i \frac{dc_{A}(t)}{dt} = \left.\left<A|V_{\textrm{int}}(t)|B\right> e^{i(\varepsilon_{A}-\varepsilon_{B})t} c_{B}(t)\right. ,
\end{equation}
\begin{equation}
\label{time_evoln_int_Cn2}
i \frac{dc_{B}(t)}{dt} = \left.\left<B|V_{\textrm{int}}(t)|A\right> e^{i(\varepsilon_{B}-\varepsilon_{A})t} c_{A}(t)\right. ,
\end{equation}
where $V_{\textrm{int}}(t)$ denotes the off-diagonal perturbative interaction in (\ref{Ham_new}).
We apply the slow turn-on perturbative method \cite{Sakurai}, in which we multiply the off-diagonal perturbative interaction in (\ref{Ham_new}) by the factor $e^{\eta t}$, where $\eta > 0$. Solving Eq.~(\ref{time_evoln_int_Cn2}) for $c_B(t)$ with the initial conditions $c_{A}(-\infty)=1$, $c_{B}(-\infty)=0$ under the assumption that $c_A(t) \approx 1$ (which is equivalent to the application of first-order time-dependent perturbation theory (TDPT)), then letting $\eta \to 0^{+}$ at the end of the calculation, gives the perturbed wavefunction corresponding to the unperturbed parity eigenstate $\left| A \right>$ to be
\begin{widetext}
\begin{equation}
\label{A_tilde_general}
\left|\tilde{A}(t)\right> = e^{-i \varepsilon_{A} t} \left\{ \left|A\right> +  \frac{H_A}{(\eps_B - \eps_A)^2 - m_{\mathrm{a}}^2} \left[m_{\mathrm{a}} \sin\left(m_{\mathrm{a}} t  \right) + i (\varepsilon_{B}-\varepsilon_{A}) \cos\left(m_{\mathrm{a}} t  \right) \right] \left|B\right> \right\} .
\end{equation}
\end{widetext}
From (\ref{A_tilde_general}), it follows that
\begin{equation}
\label{A_tilde_pnc}
\left|\tilde{A}(t)\right> = e^{-i \varepsilon_{A} t} \left[ \left|A\right> + \frac{i H_{A} \cos\left(m_{\mathrm{a}} t  \right)}{\varepsilon_{B}-\varepsilon_{A}} \left|B\right> \right] ,
\end{equation}
when $m_{\mathrm{a}} \ll \left| \varepsilon_{B}-\varepsilon_{A} \right|$, and
\begin{equation}
\label{A_tilde_edm}
\left|\tilde{A}(t)\right> = e^{-i \varepsilon_{A} t} \left[ \left|A\right> - \frac{H_{A} \sin\left(m_{\mathrm{a}} t  \right)}{m_{\mathrm{a}}} \left|B\right> \right] ,
\end{equation}
when $m_{\mathrm{a}} \gg \left| \varepsilon_{B}-\varepsilon_{A} \right|$.
Likewise, solving Eq.~(\ref{time_evoln_int_Cn}) for $c_A(t)$ with the initial conditions $c_{B}(-\infty)=1$, $c_{A}(-\infty)=0$ under the assumption that $c_B(t) \approx 1$, then letting $\eta \to 0^{+}$ at the end of the calculation, gives the perturbed wavefunction corresponding to the unperturbed parity eigenstate $\left| B \right>$ as
\begin{widetext}
\begin{equation}
\label{B_tilde_general}
\left|\tilde{B}(t)\right> = e^{-i \varepsilon_{B} t} \left\{ \left|B\right> +  \frac{H_A}{(\eps_B - \eps_A)^2 - m_{\mathrm{a}}^2}  \left[-m_{\mathrm{a}} \sin\left(m_{\mathrm{a}} t  \right) + i (\varepsilon_{B}-\varepsilon_{A}) \cos\left(m_{\mathrm{a}} t  \right) \right] \left|A\right> \right\} .
\end{equation}
\end{widetext}
From (\ref{B_tilde_general}), it follows that
\begin{equation}
\label{B_tilde_pnc}
\left|\tilde{B}(t)\right> = e^{-i \varepsilon_{B} t} \left[ \left|B\right> + \frac{i H_{A} \cos\left(m_{\mathrm{a}} t  \right)}{\varepsilon_{B}-\varepsilon_{A}} \left|A\right> \right] ,
\end{equation}
when $m_{\mathrm{a}} \ll \left| \varepsilon_{B}-\varepsilon_{A} \right|$, and
\begin{equation}
\label{B_tilde_edm}
\left|\tilde{B}(t)\right> = e^{-i \varepsilon_{B} t} \left[ \left|B\right> + \frac{H_{A} \sin\left(m_{\mathrm{a}} t  \right)}{m_{\mathrm{a}}} \left|A\right> \right] ,
\end{equation}
when $m_{\mathrm{a}} \gg \left| \varepsilon_{B}-\varepsilon_{A} \right|$. Note in particular the sign differences in the coefficients of admixture in Eqs.~(\ref{A_tilde_edm}) and (\ref{B_tilde_edm}).

From Eqs.~(\ref{A_tilde_general}) and (\ref{B_tilde_general}), we see that there exists an oscillatory PNC effect due to the purely imaginary coefficients of admixture (the purely imaginary coefficients of admixture for the opposite-parity states ensures that there are no contributions to the EDMs of the perturbed states from these coefficients), as well as an oscillatory EDM (in addition to the inherent PNC effect) due to the real coefficients of admixture. 
From Eqs.~(\ref{A_tilde_pnc}) and (\ref{B_tilde_pnc}), we see that, when $m_{\mathrm{a}} \ll \left| \varepsilon_{B}-\varepsilon_{A} \right|$, the oscillatory PNC effect dominates, while from Eqs.~(\ref{A_tilde_edm}) and (\ref{B_tilde_edm}), we see that, when $m_{\mathrm{a}} \gg \left| \varepsilon_{B}-\varepsilon_{A} \right|$, the oscillatory EDM effect dominates. 

We now consider a full calculation of the mixing of opposite-parity states caused by the interaction (\ref{H_int_NRL_temp1}). Consider the unperturbed eigenstate $\left|B\right>$, for instance. In principle, any parity eigenstate of opposite parity to that of $\left|B\right>$ can mix with the unperturbed eigenstate $\left|B\right>$. By analogy with Eq.~(\ref{B_tilde_general}), application of first-order TDPT, with account of all possible states that can mix with $\left| B \right>$, yields the following perturbed wavefunction corresponding to the unperturbed parity eigenstate $\left| B \right>$:
\begin{widetext}
\begin{equation}
\label{B_tilde_general_full}
\left|\tilde{B}(t)\right> = e^{-i \varepsilon_{B} t} \left\{ \left|B\right> + \sum_{m} \frac{H_m}{(\eps_B - \eps_m)^2 - m_{\mathrm{a}}^2}  \left[-m_{\mathrm{a}} \sin\left(m_{\mathrm{a}} t  \right) + i (\varepsilon_{B}-\varepsilon_{m}) \cos\left(m_{\mathrm{a}} t  \right) \right] \left|m\right> \right\} ,
\end{equation}
\end{widetext}
where 
\begin{equation}
\label{H_m}
H_{m} = \frac{a_{0} m_{\mathrm{a}} \left(\varepsilon_m - \varepsilon_B \right) \left< m \right| \mathbf{r}_{\mathrm{e}} \cdot \mathbf{\sigma}_{\mathrm{e}} \left| B \right>}{ f_{\textrm{a}} } ,
\end{equation}
and we have made use of the fact that $\left< m \right| \mathbf{r}_{\mathrm{e}} \cdot \mathbf{\sigma}_{\mathrm{e}} \left| B \right> =0$ if the parity eigenstate $\left| m \right>$ has the same parity as $\left| B \right>$ does, so that the sum over $m$ in (\ref{B_tilde_general_full}) runs over the complete set of unperturbed parity eigenstates for the system of interest.
Fortunately, formula (\ref{B_tilde_general_full}) can simplify tremendously depending on the system and property of the perturbed wavefunction of interest. Recalling that the axion mass $m_{\mathrm{a}}$ at present is generally believed to lie in the range $10^{-6} - 1$ eV, in some systems the condition $|\varepsilon_{B}-\varepsilon_{m}| \gg m_{\mathrm{a}}$ may hold for all states $\left| m \right>$ with parity opposite to that of $\left| B \right>$. In such a case, formula (\ref{B_tilde_general_full}) simplifies to
\begin{equation}
\label{B_tilde_general_full_simplified}
\left|\tilde{B}(t)\right> = e^{-i \varepsilon_{B} t} \left\{ \left|B\right> - \frac{i a_0 m_{\mathrm{a}} \cos\left(m_{\mathrm{a}} t  \right)}{f_{\mathrm{a}}}  \mathbf{r}_{\mathrm{e}} \cdot \mathbf{\sigma}_{\mathrm{e}} \left|B\right> \right\} ,
\end{equation}
where we have neglected the real contribution to the coefficients of admixture, which are supressed compared with the purely imaginary contribution in this case. Here $\mathbf{r}_{\mathrm{e}} \cdot \mathbf{\sigma}_{\mathrm{e}} \left|B\right>$ gives a projection onto the subspace of parity eigenstates with opposite parity to that of $\left|B\right>$. 

Note that formula (\ref{B_tilde_general_full_simplified}) also applies to nuclei, under the same assumptions made for atomic and molecular systems. Moreover, (\ref{B_tilde_general_full_simplified}) has the same form as that of the wavefunction, which gives rise to the nuclear anapole moment and reads as follows in the coordinate-space representation \cite{Khrip84}:
\begin{equation}
\label{anapole_WF}
\psi \left(\mathbf{r} \right) = \left[1 - \frac{i G_{F} g_{N} \rho_0}{\sqrt{2}} \mathbf{\sigma} \cdot \mathbf{r} \right] \psi_0 \left(\mathbf{r} \right) ,
\end{equation}
where $G_F$ is the Fermi constant of the weak interaction, $g_{N}$ is a dimensionless constant that is expressed through constants of the weak meson-nucleon interaction and is different for a proton and neutron, and $\rho_0$ is the average nuclear density. The wavefunction (\ref{anapole_WF}) gives rise to the following anapole moment \cite{Khrip84}:
\begin{equation}
\label{anapole_std}
\mathbf{a} = \frac{G_{F} g_{N} \rho_0}{\sqrt{2}} \frac{2 \pi e \mu_{N}}{m_{N}} \frac{K \mathbf{I}}{I(I+1)} \left< r^2 \right> ,
\end{equation}
where $\mathbf{I}$ is the nuclear spin, $\mu_N$ is the nucleon magnetic moment in nuclear magnetons, $m_N$ is the nucleon mass, $K=(I+1/2)(-1)^{I+1/2-l}$, with $l$ being the orbital angular momentum of the nucleon, and $\left<  r^2 \right>$ is the square radius of the nucleon of interest. Likewise, Eq.~(\ref{B_tilde_general_full_simplified}) gives rise to oscillating anapole moments associated with both the  electrons and the nucleons, which by analogy with Eqs.~(\ref{anapole_WF}) and (\ref{anapole_std}) can be written as
\begin{equation}
\label{anapole_osc_ADM}
\mathbf{a}_\lambda = \frac{a_0 m_{\mathrm{a}} \cos\left( m_{\mathrm{a}} t \right)}{f_{\mathrm{a}}} \frac{2 \pi e \mu_{\lambda}}{m_{\lambda}} \frac{K_\lambda \mathbf{I}_\lambda}{I_\lambda(I_\lambda+1)} \left< r^2 \right>_{\lambda} ,
\end{equation}
where $\lambda = e$ denotes an electron and $\lambda = N$ denotes a nucleon, $\mathbf{I}_\lambda$ denotes either the electron or nucleon spin as appropriate, and all other variables are defined analagously, depending on whether $\lambda = e$ or $\lambda = N$, to those in formula (\ref{anapole_std});  $\mu_e=1$,  $\mu_p=2.8$,  $\mu_n=-1.9$. The anapole moments induce PNC effects in atoms and molecules \cite{Khrip84}.

Interaction (\ref{H_int_NRL_temp1}) may also generate a PNC electric dipole amplitude $E_{PNC}$ between states of the same parity, e.g.~in the $6s-7s$ transition in atomic caesium, where the most accurate measurements and calculations of $E_{PNC}$ generated by the weak interaction have been performed. However, for the wavefunction (\ref{B_tilde_general_full_simplified}), obtained in the non-relativistic approximation, this amplitude vanishes and one should perform the relativistic calculation instead. Numerical relativistic many-body calculations of $E_{PNC}$ values will be presented in a separate publication.

Now suppose that we are interested solely in measuring the EDM of a state with a single valence electron in the $s$-wave. In this case, the oscillating EDM associated with the state $\left|\tilde{B}(t)\right>$, which follows from formula (\ref{B_tilde_general_full}), is
\begin{align}
\label{A_tilde_edm_expr_new}
&d_{\textrm{a}} = \left<\tilde{B}(t)\right| e \left(\mathbf{r}_{\mathrm{e}} \right)_{\textrm{z}} \left|\tilde{B}(t)\right> \notag \\
&= -\frac{3 a_{0} m_{\textrm{a}}^2 \alpha_{zz} (m_{\textrm{a}})  }{f_{\textrm{a}} \alpha} e \sin\left(m_{\mathrm{a}} t \right) ,
\end{align}
where $\alpha_{zz} (m_{\textrm{a}})$ is the dynamic polarisability with the applied frequency given by $\omega = m_{\textrm{a}}$ \cite{Atkins}:
\begin{equation}
\label{dyn_pol_m-a}
\alpha_{zz} (m_{\textrm{a}}) = 2 \sum_{m \ne B} \frac{(\eps_m - \eps_B) \left|\left<m| e \left(\mathbf{r}_{\mathrm{e}} \right)_{\textrm{z}}|B \right>\right|^2}{(\eps_B - \eps_m)^2 - m_{\mathrm{a}}^2} .
\end{equation}
As an application of formula (\ref{A_tilde_edm_expr_new}), we consider hydrogen and the alkali metals in their respective ground states. The polarisabilty of an alkali atom is dominated by its valence $s$-wave electron. If we assume that the condition $|\varepsilon_{B}-\varepsilon_{m}| \gg m_{\mathrm{a}}$ holds for all states $\left| m \right>$ with parity opposite to that of $\left| B \right>$, then $\alpha_{zz} (m_{\textrm{a}}) \approx \alpha_{zz} (0)$. Furthermore, since the ground states of interest are spherically symmetric, $\alpha_{zz} (0) = \alpha_s$, where $\alpha_s$ is the scalar  static polarisability. Thus (\ref{A_tilde_edm_expr_new}) becomes
\begin{equation}
\label{A_tilde_edm_expr_new2}
d_{\textrm{a}} = -\frac{3 a_{0} m_{\textrm{a}}^2 \alpha_{s} }{f_{\textrm{a}} \alpha} e \sin\left(m_{\mathrm{a}} t \right) .
\end{equation}
We further assume that axions saturate the entire CDM content of the universe and that there is no fine tuning of the misalignment angle $\theta_{\textrm{i}}$ (that is, $\theta_{\textrm{i}} \sim 1$). We thus take $m_{\textrm{a}} \sim 10^{-4}$ eV and $f_{\textrm{a}} \sim 10^{20}$ eV for our estimate \cite{PDG12}. In order to ascertain an estimate for $a_{0}$, we take the non-relativistic limit of Eq.~(\ref{energy_density}) and solve
\begin{equation}
\label{a_0}
\frac{a_{0}^{2} m_{\mathrm{a}}^{2}}{2} = \rho_{\textrm{CDM}} ,
\end{equation}
where $\rho_{\textrm{CDM}} \sim 7.6 \cdot 10^{-4} \textrm{eV}^{4}$ ($0.4~\textrm{GeV}/\textrm{cm}^{3}$ in more conventional units) is the local CDM density \cite{Catena10,Salucci10,Pato10,Catena12,Kim_RMP_10,PDG12}. This gives $a_{0} \sim 4 \cdot 10^{2}$ eV. We summarise our estimates of the induced EDMs for the group I elements according to Eq.~(\ref{A_tilde_edm_expr_new2}) in Table \ref{table:alkalis_edm}, from which we see that the oscillating EDM induced in caesium is of the same order-of-magnitude as our estimate for the oscillating EDM induced in the atomic species $^{199}$Hg through hadronic mechanisms (see Eq.~(\ref{d_Hg_total})). This is significant, because $^{199}$Hg currently provides the most sensitive probe for static EDM measurements in diamagnetic atomic species \cite{Romalis09,Romalis13A}. Note, however, that when the condition $|\varepsilon_{B}-\varepsilon_{m}| \gg m_{\mathrm{a}}$ holds for all states $\left| m \right>$ with parity opposite to that of $\left| B \right>$, the EDM effect is suppressed compared with the PNC effect by a factor $\sim \frac{m_{\textrm{a}}}{\Delta \eps} \ll 1$, as evident from Eq.~(\ref{B_tilde_general_full}). (See also Eqs.~(\ref{A_tilde_edm_expr}) and (\ref{A_tilde_edm_estimate}) for the other limiting case.)

\begin{table}[h!]
\caption{Estimated magnitudes of oscillating EDMs induced in Group I atomic species, as predicted by formula (\ref{A_tilde_edm_expr_new2}). Values for scalar static polarisabilities were taken from Ref.~\cite{table_static_pol} and references therein.}
\begin{center}
\begin{tabular}{|c|c|}
\hline
Species &  $\left| d_\textrm{a} \right| / 10^{-40}$ e $\cdot$ cm~$\sin\left(m_{\mathrm{a}} t \right)$  \\
\hline
H & 7 \\
\hline
Li & 300 \\
\hline
Na & 300 \\
\hline
K & 500 \\
\hline
Rb & 500 \\
\hline
Cs & 600 \\
\hline
\end{tabular}
\label{table:alkalis_edm}
\end{center}
\end{table}

If one is again interested solely in measuring the EDM of a state, then, recalling Eq.~(\ref{B_tilde_edm}), it might happen that the condition $m_{\mathrm{a}} \gg \left| \varepsilon_{B}-\varepsilon_{m} \right|$ might only be satisfied for one or possible only a few states $\left| m \right>$ of opposite parity to that of $\left| B \right>$, in which case only these states will contribute significantly to the overall EDM. The existence of a pair of nearly degenerate levels of opposite parity in atomic and molecular systems is quite uncommon, so the condition $m_{\mathrm{a}} \gg \left| \varepsilon_{B}-\varepsilon_{m} \right|$ is most likely to be satisfied for either one or no such state $\left| m \right>$. Since we are not interested in the imaginary component of the coefficients of admixture in (\ref{B_tilde_general_full}), the perturbed wavefunction for the purpose of calculating the EDM of a state takes on the same form as (\ref{B_tilde_edm}) if the condition $m_{\mathrm{a}} \gg \left| \varepsilon_{B}-\varepsilon_{m} \right|$ holds for only the one pair of states, and an analagous form with additional admixture terms for each additional state $\left| m \right>$ that satisfies the condition $m_{\mathrm{a}} \gg \left| \varepsilon_{B}-\varepsilon_{m} \right|$. Now suppose that the condition $m_{\mathrm{a}} \gg \left| \varepsilon_{B}-\varepsilon_{m} \right|$ holds only for the state $\left| m \right>$ with $m=A$ and that $m_{\mathrm{a}} \ll \left| \varepsilon_{B}-\varepsilon_{m} \right|$ for all other states with opposite parity to that of $\left| B \right>$. In this case, from Eq.~(\ref{B_tilde_edm}), the oscillating EDM associated with the state $\left|\tilde{B}(t)\right>$ is given by
\begin{equation}
\label{A_tilde_edm_expr}
d_{\textrm{a}} =  \frac{2a_{0} \left(\varepsilon_{A}-\varepsilon_{B} \right) \left< A \right| \mathbf{r}_{\mathrm{e}} \cdot \mathbf{\sigma}_{\mathrm{e}} \left| B \right> \left< B \right| \left(\mathbf{r}_{\mathrm{e}} \right)_{\textrm{z}} \left| A \right> }{f_{\textrm{a}}} e \sin\left(m_{\mathrm{a}} t \right) .
\end{equation}
The matrix elements of interest in (\ref{A_tilde_edm_expr}) are typically $\left|\left< B \right| \left(\mathbf{r}_{\mathrm{e}} \right)_{\textrm{z}} \left| A \right> \right| \sim \left| \left< A \right| \mathbf{r}_{\mathrm{e}} \cdot \mathbf{\sigma}_{\mathrm{e}} \left| B \right> \right| \sim a_{\textrm{B}}$, where $a_{\textrm{B}} = \frac{1}{\alpha m_{e} }$ is the Bohr radius. Taking $\left|\varepsilon_{A}-\varepsilon_{B} \right| = 10^{-5}$ eV and using our previous estimates for the other quantities in Eq.~(\ref{A_tilde_edm_expr}), an order-of-magnitude estimate of the oscillating EDM induced in an atomic or molecular species, possessing a single nearly degenerate pair of opposite-parity states, by a background axion field via the interaction (\ref{L_int}) is 
\begin{equation}
\label{A_tilde_edm_estimate}
\left|d_{\textrm{a}}\right| \sim 1 \cdot 10^{-34}~\mathrm{e \cdot cm}~\sin\left(m_{\mathrm{a}} t \right) ,
\end{equation}
which is of the same order-of-magnitude as estimates for the oscillating EDM of a free neutron, which arises from the coupling of the axion field to gluons, in Refs.~\cite{Graham11,Graham13} (see also Eq.~(\ref{d_n-theta_est}) later in this work), and is roughly two orders-of-magnitude greater than our estimates for the oscillating EDM induced in caesium through the same mechanism (see Table \ref{table:alkalis_edm}) and the atomic species $^{199}$Hg through hadronic mechanisms (see Eq.~(\ref{d_Hg_total})). Note that closed shell electronic and nucleonic configurations do not contribute to the overall atomic or molecular EDM, since any EDM must be directed along the total angular momentum of the system and the total angular momentum of such closed shell configurations is necessarily zero. The possibility of an EDM of any sort arising from the interaction (\ref{L_int}) is quite intriguing, since in the static limit ($\cos\left(m_{\mathrm{a}} t  \right) \to$ constant), the Hamiltonian (\ref{Ham_new}) cannot give rise to an EDM - see the wavefunctions (\ref{A_tilde_pnc}) and (\ref{B_tilde_pnc}) for comparison.

An oscillatory EDM can be detected through energy level shifts or one of the methods suggested in Refs.~\cite{Graham11,Graham13}. We also describe a general scheme, which is based on the Stark-interference technique, for the detection of an oscillatory EDM and oscillatory PNC effects in Sec.~\ref{Sec:pnc}.

Finally, we note that if one solves Eq.~(\ref{time_evoln_int_Cn}) for $c_A(t)$ with the initial conditions $c_{B}(-\infty)=1$, $c_{A}(-\infty)=0$ under the assumption that $c_B(t) \approx 1$, then one finds that the magnitude of the coefficient of admixture for the parity eigenstate $\left| A \right>$ in the perturbed wavefunction corresponding to the unperturbed parity eigenstate $\left| B \right>$ tends to infinity as $m_{\mathrm{a}} \to \left| \varepsilon_{B}-\varepsilon_{A} \right|$, as evident from Eq.~(\ref{B_tilde_general}). The condition $m_{\mathrm{a}} = \left| \varepsilon_{B}-\varepsilon_{A} \right|$ indicates that a resonance transition is being induced between the parity eigenstates $\left| A \right>$ and $\left| B \right>$. The singularity in the wavefunction is an artefact of our assumption of first-order TDPT and neglect of the natural widths of the states considered, at least one of which must be non-zero. This artificial singularity is removed when we take into account the natural widths of the states of interest, but the resonant behaviour remains, as we now show. Suppose that $\left| \eps_B - \eps_A \right| \gg \Gamma_A / 2 \gg \Gamma_B /2$. In the presence of the off-diagonal interaction (\ref{M_elt_5}), the Hamiltonian for the two-level subspace spanned by the $\left| A \right>$ and $\left| B \right>$ parity eigenstates, with account of the more dominant width of the two only, reads
\begin{equation}
\label{Ham_new_width}
H \left(t \right) =
\left[ \begin{array}{cc}
\varepsilon_{A} - i\Gamma_A /2 & iH_{A} \cos\left(m_{\mathrm{a}} t  \right) \\
-iH_{A} \cos\left(m_{\mathrm{a}} t  \right) & \varepsilon_{B} 
\end{array} \right] . 
\end{equation}
We write the unperturbed system wavefunction projection onto the two-level subspace of interest, again with account of the more dominant width of the two only, as
\begin{equation}
\label{WF_width}
\left| \psi (t)\right> = c_{A}(t) e^{-i \varepsilon_{A}t} e^{-\Gamma_A t /2} \left| A \right> + c_{B}(t) e^{-i \varepsilon_{B}t} \left| B \right> ,
\end{equation}
from which follows the following coupled differential equation:
\begin{equation}
\label{time_evoln_int_Cn_width}
i \frac{dc_{A}(t)}{dt} = \left.\left<A|V_{\textrm{int}}(t)|B\right> e^{i(\varepsilon_{A}-\varepsilon_{B})t} e^{+\Gamma_{A} t /2} c_{B}(t)\right. ,
\end{equation}
where $V_{\textrm{int}}(t)$ is the off-diagonal perturbation in (\ref{Ham_new_width}).
Suppose without of loss generality that $m_{\mathrm{a}} = \varepsilon_{A}-\varepsilon_{B}$. Solving Eq.~(\ref{time_evoln_int_Cn_width}) for $c_A(t)$ with the initial conditions $c_{B}(-\infty)=1$, $c_{A}(-\infty)=0$ under the assumption that $c_B(t) \approx 1$, gives the perturbed wavefunction corresponding to the unperturbed parity eigenstate $\left| B \right>$ as
\begin{equation}
\label{B_tilde_resn}
\left|\tilde{B}(t)\right> = e^{-i \varepsilon_{B} t} \left[ \left|B\right> + \frac{H_{A} e^{-i m_{\mathrm{a}} t}  }{\Gamma_{A}} \left|A\right> \right] ,
\end{equation}
where we have ignored contributions from all the other states, which may be admixed into (\ref{B_tilde_resn}), since their coefficients of admixture are likely to be overwhelming small compared with that for the parity eigenstate $\left|A\right>$. Comparing (\ref{B_tilde_resn}) with (\ref{B_tilde_pnc}) and (\ref{B_tilde_edm}), we see that when the resonant condition $m_{\mathrm{a}} = \left| \varepsilon_{B}-\varepsilon_{A} \right|$ is satisfied, the coefficient of admixture is enhanced by a factor of $\sim \frac{m_{\textrm{a}}}{\Gamma_A}$ compared with the values away from the resonance. Enhancement is greater when the natural widths of the states are smaller. Molecular species are particularly advantageous in this regard, with widths of $\sim 1$ Hz quite common (see e.g.~Ref.~\cite{Flambaum_FC_07}). For the values $m_{\textrm{a}} = 10^{-4}$ eV and $\Gamma_A = 4 \cdot 10^{-15}$ eV, the enhancement on resonance is ten orders-of-magnitude from this consideration alone.


\section{Overview of conventional Stark-interference experiments}
\label{Sec:Overview}
\begin{figure}[h!]
\begin{center}
\includegraphics[width=8.5cm]{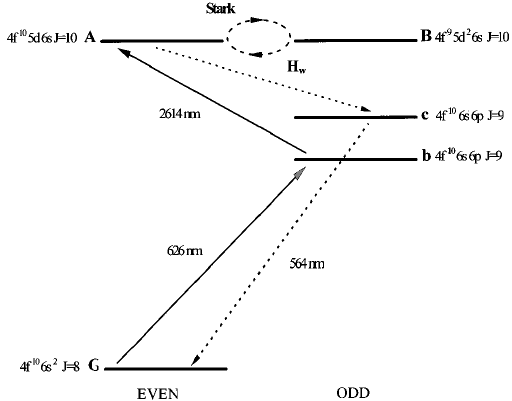}
\caption{Schematic of relevant parity eigenstates in atomic dysprosium experiment. Figure reproduced from Ref.~\cite{Nguyen97} with permission.}
\label{fig:Dy_levels}
\end{center}
\end{figure}

We recapitulate the essence of one particular variant of the Stark-interference technique through recourse to the atomic dysprosium experiment of Ref.~\cite{Nguyen97}, which seeks to measure the static weak interaction-induced mixing of the pair of opposite parity eigenstates, $\left| A \right>$ and $\left| B \right>$, which are first brought to near degeneracy through the application of a uniform magnetic field (Fig.~\ref{fig:Dy_levels}). Note that in this section, $\left| A \right>$ and $\left| B \right>$ are specific states, whereas in the other sections of this paper, they are arbitrary. State $\left| A \right>$ is the more unstable state of the two, with a lifetime of $\tau_{A} = 7.9$ $\mathrm{\mu}$s, while state $\left| B \right>$ is essentially stable on the time scale of the whole experiment, with a lifetime in excess of $200$ $\mathrm{\mu}$s. The parity eigenstate $\left| A \right>$ is populated by a broad (in frequency) two-step pulse excitation from the ground state $\left| G \right>$ via state $\left|b\right>$. This is followed almost immediately by a $\pi$-pulse transfer from $\left| A \right>$ to $\left| B \right>$. In order to obtain clean initial conditions for the subsequent Stark-interference step, a waiting period of roughly $10~\tau_{A}$ is executed. During this time, the $\left| A \right>$ component of the wavefunction for the system undergoes decay, leaving $\left| B \right>$ as the only occupied parity eigenstate of the two-level system spanned by $\left| A \right>$ and $\left| B \right>$. Thus the initial conditions for the Stark-interference step are: $c_{B}(0) = 1$, $c_{A}(0) = 0$. The Stark-interference step consists of applying an oscillating electric field of the form $E\left(t\right) = E_{0}\cos\left(\omega t \right)$, which induces oscillations in the population of state $\left| A \right>$, with the frequency of the applied electric field being much larger than the energy separation, $\Delta = \varepsilon_{B} - \varepsilon_{A}$, between states $\left| A \right>$ and $\left| B \right>$. The time-dependent Hamiltonian describing the system in the presence of the weak interaction and oscillating electric field reads
\begin{equation}
\label{Ham}
H \left(t \right) =
\left[ \begin{array}{cc}
\varepsilon_{A} - i \Gamma_{A}/2 & iH_{w} + dE_{0} \cos\left(\omega t\right) \\
-iH_{w} + dE_{0} \cos\left(\omega t\right) & \varepsilon_{B} - i \Gamma_{B}/2 
\end{array} \right] , 
\end{equation}
where the purely imaginary, time-reversal invariant weak interaction matrix elements between the states $\left|A\right>$ and $\left|B\right>$ are defined by $V_{AB}^{(W)} = i H_{W} = -V_{BA}^{(W)}$, $d$ denotes the real electric dipole matrix element between the states $\left|A\right>$ and $\left|B\right>$, and $\Gamma_{A}$ and $\Gamma_{B}$ are the natural widths of the states $\left|A\right>$ and $\left|B\right>$ respectively. The corresponding observable, which is derived from TDPT, reads as follows (with $\Gamma_{B} = 0$):
\lineskiplimit=0pt\relax
\begin{align}
\label{old_Dy_formula}
\left|\left<A\left|\right.\psi\left(t\right)\right>\right|^{2}  &= \left(\frac{dE_{0}}{\omega}\right)^{2} \sin^{2}\left(\omega t\right) \notag \\ 
&- \frac{2dE_{0}H_{w}}{\omega} \left(\frac{\Delta}{\Delta^2+\Gamma_{A}^{2}/4}\right) \sin\left(\omega t\right) .
\end{align}
Measurement of the characteristic second term in Eq.~(\ref{old_Dy_formula}), which changes sign upon the reversal of the applied electric field direction (the first term in (\ref{old_Dy_formula}) does not change sign) and has a different time dependence compared with the first term, then permits a determination of the the magnitude of the weak interaction matrix element $\left|H_{W}\right|$.

\section{Detection of oscillating EDMs and oscillating PNC effects using Stark-interference technique}
\label{Sec:pnc}

Suppose that $\Gamma_A / 2 \gg \Gamma_B /2$, and that the parity eigenstate $\left| B \right>$ has been populated, with the opposite-parity eigenstate $\left| A \right>$ unpopulated, say, through an E1 transition from a remote state, induced by a laser pulse of sufficiently short duration that the frequency-width of the pulse ($\Gamma_{p}$) is much greater than the energy separation between the parity eigenstates $\left| A \right>$ and $\left| B \right>$, that is, $\Gamma_{p} \gg \left|\eps_{\textrm{B}} - \eps_{\textrm{A}} \right|$. At time $t=0$, we apply a static electric field $\mathbf{E} = E_{0} \mathbf{\hat{z}}$ to the system. The Hamiltonian for the two-level subspace spanned by the $\left| A \right>$ and $\left| B \right>$ parity eigenstates reads, with the aid of Eq.~(\ref{M_elt_5})
\begin{equation}
\label{Ham_new_width_Stark}
H \left(t \right) =
\left[ \begin{array}{cc}
\varepsilon_{A} - i\Gamma_A /2 & iH_{A} \cos\left(m_{\mathrm{a}} t  \right) +dE_{0} \\
-iH_{A} \cos\left(m_{\mathrm{a}} t  \right) +dE_{0} & \varepsilon_{B} 
\end{array} \right] ,
\end{equation}
where $d$ is the real electric dipole matrix element between the $\left| A \right>$ and $\left| B \right>$ parity eigenstates, and we have taken into account the more dominant width of the two only. For sufficiently small perturbations, $c_{B}(t) \approx 1$ and we find by solving the differential equation (\ref{time_evoln_int_Cn_width}) subject to the initial conditions $c_{B}(0)=1$, $c_{A}(0)=0$
\begin{widetext}
\begin{equation}
\label{}
c_A (t) = +\frac{dE_{0} \left(e^{-i  \Delta t} e^{\Gamma_A t/2} -1 \right)}{\Delta +i\Gamma_A / 2} + \frac{H_A}{2i} \left\{ \frac{\left[e^{i (m_{\mathrm{a}} - \Delta) t} e^{\Gamma_A t/2} -1 \right]}{m_{\mathrm{a}} - \Delta -i \Gamma_A / 2} - \frac{\left[e^{-i (m_{\mathrm{a}} + \Delta) t} e^{\Gamma_A t/2} -1 \right]}{m_{\mathrm{a}} + \Delta  + i\Gamma_A / 2} \right\} ,
\end{equation}
where $\Delta = \eps_B - \eps_A$. If we detect the parity eigenstate $\left| A \right>$, then the predicted observable for times $t$ such that $\Gamma_A t / 2 \gg 1$ and to first order in $H_A$ is given by
\begin{align}
\label{NP,NP_obs_new_resn}
&\left|\left< A \left|\right.\psi\left(t\right)\right>\right|^{2}  =   \frac{\left(dE_{0}\right)^{2} }{\Delta^2 + \Gamma_A^2 /4} \notag \\
&- H_A dE_0 \left[ \frac{\frac{-\Gamma_A m_{\textrm{a}}}{2} \cos(m_{\textrm{a}} t) + (\Delta^2 +\Gamma_A^2 / 4 - m_{\textrm{a}} \Delta) \sin(m_{\textrm{a}} t)}{(\Delta^2 +\Gamma_A^2 / 4 - m_{\textrm{a}} \Delta)^2 + \Gamma_{A}^2 m_{\textrm{a}}^2 / 4} + \frac{\frac{\Gamma_A m_{\textrm{a}}}{2} \cos(m_{\textrm{a}} t) - (\Delta^2 +\Gamma_A^2 / 4 + m_{\textrm{a}} \Delta) \sin(m_{\textrm{a}} t)}{(\Delta^2 +\Gamma_A^2 / 4 + m_{\textrm{a}} \Delta)^2 + \Gamma_{A}^2 m_{\textrm{a}}^2 / 4}  \right]  .
\end{align}
\end{widetext}
The second term in Eq.~(\ref{NP,NP_obs_new_resn}) is the term of interest - it is distinguished from the first term by the presence of the time-dependent factors $\cos(m_{\textrm{a}} t)$ and $\sin(m_{\textrm{a}} t)$, as well as its sign reversal upon the reversal of the applied electric field direction ($E_{0} \to -E_{0}$); the first term, however, is unchanged upon the reversal of the applied electric field direction. The measurement of this second term provides a means of determining the axion parameters $m_{\textrm{a}}$ and $a_{0}/f_{\textrm{a}}$. We note also that the presence of a static weak interaction between the two states of interest, characterised by a matrix element of magnitude $\left|H_{w} \right|$ (see e.g.~the Hamiltonian (\ref{Ham})), cannot give rise to an analogous observable term that both changes sign upon the reversal of the applied electric field direction and contains either of the time-dependent factors $\cos(m_{\textrm{a}} t)$ or $\sin(m_{\textrm{a}} t)$. Hence the second term in Eq.~(\ref{NP,NP_obs_new_resn}) bears a unique signature.

\section{Axion-induced oscillating EDMs generated through hadronic mechanisms}
\label{Sec:hadronic}
For a neutral, non-relativistic classical or quantum system that consists of pointlike, charged particles, which possess permanent EDMs and interact with each other only by means of the electrostatic interaction, there exists complete shielding of the constituent EDMs when the system is exposed to an arbitary external electric field. This is the essence of Schiff's theorem \cite{Schiff63}. In real atomic systems, shielding is incomplete and so a permanent atomic EDM can in principle be borne. For heavy atoms, such as $^{199}$Hg and $^{225}$Ra, the primary cause of incomplete screening is finite nuclear size and the degree of incomplete screening is quantified by the nuclear Schiff moment $S$. It is common to express the nuclear contribution to the EDM of an atom in terms of $S$. Calculations have been performed to determine how the nuclear Schiff moment-induced EDMs of $^{199}$Hg (see e.g.~\cite{Flambaum85,Dzuba09,Latha09,Flambaum86,Dzuba02}) and $^{225}$Ra (see e.g.~\cite{Dzuba02,Spevak96,Spevak97}) depend on $S$. In the present work, we use the result of Ref.~\cite{Dzuba09} for $^{199}$Hg
\begin{equation}
\label{d-S-Hg}
d \left(^{199} \textrm{Hg} \right) = -2.6 \cdot 10^{-17} \left( \frac{S}{\textrm{e} \cdot \textrm{fm} ^{3}} \right) \textrm{e} \cdot \textrm{cm} ,
\end{equation}
and the result of Ref.~\cite{Dzuba02} for $^{225}$Ra
\begin{equation}
\label{d-S-Ra}
d \left(^{225} \textrm{Ra} \right) = -8.5 \cdot 10^{-17} \left( \frac{S}{\textrm{e} \cdot \textrm{fm} ^{3}} \right) \textrm{e} \cdot \textrm{cm} .
\end{equation}

There are two distinct contributions to the EDMs of $^{199}$Hg and $^{225}$Ra from hadronic mechanisms. One contribution is from the P,T-violating nucleon-nucleon interaction mediated by pion exchange (see e.g.~Refs.~\cite{Dmitriev03,Haxton83,Khripl00}), which can be presented as
\begin{widetext}
\begin{equation}
\label{NN'-PT-odd_int}
W \left(\mathbf{r}_1 - \mathbf{r}_2 \right) = - \frac{g}{8 \pi m_{N}} \left[\mathbf{\nabla}_1 \left( \frac{e^{-m_{\pi}r_{12}}}{r_{12}} \right) \right] \cdot \left\{(\mathbf{\sigma}_1 - \mathbf{\sigma}_2) [\bar{g}_{0} \mathbf{\tau}_1 \cdot \mathbf{\tau}_2 + \bar{g}_{2} (\mathbf{\tau}_1 \cdot \mathbf{\tau}_2 - 3 \tau_{1z} \tau_{2z}) ] + \bar{g}_1 (\tau_{1z}\mathbf{\sigma}_1 - \tau_{2z}\mathbf{\sigma}_2)   \right\} ,
\end{equation}
\end{widetext}
where $g=13.5$ is the strong P,T-conserving $\pi N N$ coupling constant, $m_N$ is the nucleon mass, $m_\pi$ is the pion mass, $\mathbf{\sigma}$ is the nucleon spin, $\mathbf{\tau}$ is the nucleon Pauli isospin matrix in vectorised form and $r_{12}$ is the internucleon separation. The constants $\bar{g}_0$, $\bar{g}_1$ and $\bar{g}_2$ represent the strengths of the isoscalar, isovector and isotensor couplings respectively. The interaction (\ref{NN'-PT-odd_int}) gives rise a nuclear Schiff moment, which can be presented in the following form:
\begin{equation}
\label{general-S}
S = g(b_0 \bar{g}_0 + b_1 \bar{g}_1 + b_2 \bar{g}_2) ~ \textrm{e} \cdot \textrm{fm}^3 .
\end{equation}
Calculation of the parameters $b_0$, $b_1$ and $b_2$ in expression (\ref{general-S}) carries a large theoretical uncertainty and is strongly dependent on the particular phenomenological model chosen (see e.g.~Refs.~\cite{Flambaum85,Flambaum86,Dmitriev03,Dmitriev05,Jesus05,Ban10} for calculations pertaining to $^{199}$Hg and Ref.~\cite{Dob05} pertaining to $^{225}$Ra). For $^{199}$Hg, we use the results of the most recent calculation of Ref.~\cite{Ban10} for the Skyrme interaction SLy4 \cite{Chabanat98} solved in the full Hartree-Fock (projected) approximation: $b_0 = 0.013$, $b_1 = -0.006$ and $b_2 = 0.022$. For $^{225}$Ra, we use the results of Ref.~\cite{Dob05} for the Skyrme interaction SLy4: $b_0 =-3.0$, $b_1 = 16.9$ and $b_2 = -8.8$. 

So far our discussion has been general. The link of the above discussion to axion-induced effects is made when one recalls that the QCD Lagrangian contains the P,CP-violating term (see e.g.~Refs.~\cite{Haxton83,Belavin75,tHooft76,Jackiw76,Callan76})
\begin{equation}
\label{L_QCD_CP}
\mathcal{L}_{\textrm{QCD}}^{\theta} = \theta \frac{g^2}{32\pi^2} F_{a}^{\mu \nu} F_{a \mu \nu}^{*} , 
\end{equation}
where $\theta$ is the dimensionless parameter, which quantifies the degree of CP-violation, $F$ and $F^{*}$ are the gluonic field tensor and its dual respectively, $a$ is the colour index and $g^2/4\pi$ is the color coupling constant. Account of weak interaction effects results in a shift of $\theta$ from its bare value to the observable value $\bar{\theta}$ (see e.g.~Ref.~\cite{Peccei81} and references therein). The $\bar{\theta}$ term is an isoscalar and so contributes to the CP-violating isoscalar coupling constant $\bar{g}_0$ in Eq.~(\ref{NN'-PT-odd_int}) as follows \cite{Crewther79,Crewther80}:
\begin{equation}
\label{g_0-theta}
\bar{g}_0 = 0.027 \bar{\theta} .
\end{equation}
In axion models, the physically observable parameter $\bar{\theta}$ is recast into the form of an axion field, $\phi \left(\mathbf{r},t \right) / f_{\textrm{a}}$. Thus a background axion field can induce an oscillating EDM in atomic species through the P,T-violating nucleon-nucleon interaction, which for the case of $^{199}$Hg is given by
\begin{align}
\label{d_Hg_NN}
d \left(^{199} \textrm{Hg} \right) &= -1.2 \cdot 10^{-19} \frac{a_{0} }{f_{\textrm{a}}} \sin \left(m_{\textrm{a}} t \right) ~ \textrm{e} \cdot \textrm{cm} \notag \\
&= -5 \cdot 10^{-37} \sin \left(m_{\textrm{a}} t \right) ~ \textrm{e} \cdot \textrm{cm} , 
\end{align}
where we have used Eqs.~(\ref{d-S-Hg}), (\ref{general-S}) and (\ref{g_0-theta}), the known value for $g$ and calculated value for $b_0$ in $^{199}$Hg, as well as Eq.~(\ref{axion_field}) in the first line of (\ref{d_Hg_NN}), while in the second line of (\ref{d_Hg_NN}) we have used our estimate $\frac{a_0}{f_{\textrm{a}}} = 4 \cdot 10^{-18}$ from Sec.~\ref{Sec:Theory}. Likewise, the oscillating EDM induced in $^{225}$Ra by the same mechanism is given by
\begin{align}
\label{d_Ra_NN}
d \left(^{225} \textrm{Ra} \right) &= 9.3 \cdot 10^{-17} \frac{a_{0} }{f_{\textrm{a}}} \sin \left(m_{\textrm{a}} t \right) ~ \textrm{e} \cdot \textrm{cm} \notag \\
&= 4 \cdot 10^{-34} \sin \left(m_{\textrm{a}} t \right) ~ \textrm{e} \cdot \textrm{cm} , 
\end{align}
where we have used Eqs.~(\ref{d-S-Ra}), (\ref{general-S}) and (\ref{g_0-theta}), the known value for $g$ and calculated value for $b_0$ in $^{225}$Ra, as well as Eq.~(\ref{axion_field}) in the first line of (\ref{d_Ra_NN}), while in the second line of (\ref{d_Ra_NN}) we have again used our estimate $\frac{a_0}{f_{\textrm{a}}} = 4 \cdot 10^{-18}$. 

The second contribution to an atomic EDM from hadronic mechanisms arises from the intrinsic EDMs of valence nucleons within the nucleus of the atomic species of interest. $^{199}$Hg and $^{225}$Ra both possess odd-neutron, even-proton nuclei. In the single-particle approximation of the nuclear shell model \cite{Greiner}, the contribution to the Schiff moments of $^{199}$Hg and $^{225}$Ra from their valence nucleon EDMs is simply that due to the EDM of a single neutron, induced by an axion field, $d_{n}$, multiplied by the appropriate Schiff screeening factor, which is much less than unity. The dependence of $d_{n}$ on $\bar{\theta}$ is given by \cite{Ritz99}
\begin{equation}
\label{d_n-theta}
d_n = 1.2 \cdot 10^{-16} \bar{\theta} ~ \textrm{e} \cdot \textrm{cm} .
\end{equation}
With our estimate $\frac{a_0}{f_{\textrm{a}}} = 4 \cdot 10^{-18}$ and from Eq.~(\ref{d_n-theta}), our estimate for the axion-induced EDM of a free neutron is
\begin{equation}
\label{d_n-theta_est}
d_n = 5 \cdot 10^{-34} \sin \left(m_{\textrm{a}} t \right) ~ \textrm{e} \cdot \textrm{cm} .
\end{equation}
Note that our estimate (\ref{d_n-theta_est}) differs from those in Refs.~\cite{Graham11,Graham13}, due to differences in estimates for $\frac{a_0}{f_{\textrm{a}}}$. On the basis of (\ref{d_Ra_NN}) and (\ref{d_n-theta_est}), we can see that in the single-particle approximation, the contribution of valence nucleon EDMs to the EDM of $^{225}$Ra is negligible compared with the contribution from the P,T-violating nucleon-nucleon interaction. Thus Eq.~(\ref{d_Ra_NN}) is a good estimate for the axion-induced EDM of $^{225}$Ra through hadronic mechanisms. The contribution of valence nucleon EDMs to the EDM of $^{199}$Hg cannot be neglected, however. The EDM of $^{199}$Hg arising from the EDMs of its valence nucleons is due predominantly to the Schiff moment induced by the EDM of the valence neutron, with the EDMs of core protons also contributing to a lesser extent due to configuration mixing \cite{Dzuba02,Dmitriev03L}. We use the following result of Ref.~\cite{Dmitriev03L} for the Schiff moment of $^{199}$Hg induced by the EDMs of its constituent protons and neutrons:
\begin{equation}
\label{S-dn-dp_Hg}
S (^{199} \textrm{Hg}) = (1.9 d_n + 0.2 d_p) ~ \textrm{fm}^2 .
\end{equation}
If we neglect the contribution from the proton EDM in Eq.~(\ref{S-dn-dp_Hg}), then the EDM of the valence neutron, induced by an axion field, in $^{199}$Hg contributes the following amount to the axion-induced oscillating EDM of $^{199}$Hg:
\begin{align}
\label{d_Hg_n}
d \left(^{199} \textrm{Hg} \right) &= -5.9 \cdot 10^{-20} \frac{a_{0} }{f_{\textrm{a}}} \sin \left(m_{\textrm{a}} t \right) ~ \textrm{e} \cdot \textrm{cm} \notag \\
&= -2 \cdot 10^{-37} \sin \left(m_{\textrm{a}} t \right) ~ \textrm{e} \cdot \textrm{cm} , 
\end{align}
where we have used Eqs.~(\ref{d-S-Hg}), (\ref{S-dn-dp_Hg}) and (\ref{d_n-theta}), as well as Eq.~(\ref{axion_field}) in the first line of (\ref{d_Hg_n}), while in the second line of (\ref{d_Hg_n}) we have used our estimate $\frac{a_0}{f_{\textrm{a}}} = 4 \cdot 10^{-18}$ from Sec.~\ref{Sec:Theory}. From (\ref{d_Hg_NN}) and (\ref{d_Hg_n}), the overall axion-induced EDM of $^{199}$Hg through both hadronic mechanisms is hence
\begin{align}
\label{d_Hg_total}
d \left(^{199} \textrm{Hg} \right) &= -1.8 \cdot 10^{-19} \frac{a_{0} }{f_{\textrm{a}}} \sin \left(m_{\textrm{a}} t \right) ~ \textrm{e} \cdot \textrm{cm} \notag \\
&= -7 \cdot 10^{-37} \sin \left(m_{\textrm{a}} t \right) ~ \textrm{e} \cdot \textrm{cm} .
\end{align}
Comparing our estimates in (\ref{d_Ra_NN}) and (\ref{d_Hg_total}), we see that $^{225}$Ra can offer roughly a three order-of-magnitude enhancement in terms of its axion-induced oscillating EDM generated through hadronic mechanisms compared with $^{199}$Hg. This is due to both collective effects and small energy separation between members of the parity doublet of interest, which occurs in nuclei with octupolar deformation and results in a significant enhancement of the nuclear Schiff moment \cite{Spevak96,Spevak97}. Some other systems with similar enhancement of the nuclear Schiff moment through such mechanisms include $^{223}$Ra, $^{223}$Rn, $^{223}$Fr and $^{229}$Pa \cite{Spevak96,Spevak97}. These systems should also exhibit analogous enhancements in the magnitudes of their oscillating EDMs generated through the P,T-violating nucleon-nucleon interaction.

\section{Axions in gravitational fields: spin-gravity and spin-axion momentum couplings}
\label{Sec:grav}

All of the results of the previous sections, which were derived from the interaction Lagrangian density (\ref{L_int}), assumed that the interaction took place in flat, Minkowskian spacetime. However, experiments for axion detection, which involve atomic, molecular and nuclear systems, invariably take place on the surface of the Earth, where there exists a gravitational field directed radially inwards. As a result of the Earth's gravitational field (or any other such spherically symmetric gravitational field for that matter), there is an increased axion density near the surface of the Earth (or in general near the surface of the gravitating body of interest) compared with that at infinitely large distances away from the Earth, which can result in an enhancement in axion-induced effects. In this section, we focus on a relative of the axion-wind effect of Refs.~\cite{Victor13,Graham13} and show that the interaction of the spin of either an electron or nucleon, which also interacts with an axion field, with the gravitational field gradient of a gravitating body can give rise to axion-induced observable effects, which differ from the axion-wind effect - see (\ref{H_int_NRL_spat}) for the Hamiltonian responsible for the conventional axion-wind effect in flat spacetime.

The potential experienced by an axion in the Earth's gravitational field is Coulomb-like. The wavefunction of an axion in a continuum state of the Earth's gravitational field, propagating toward the Earth along the $z$-axis with momentum $\mathbf{p}_{\mathrm{a}} = p_{\textrm{a}} \mathbf{\hat{z}}$, hence reads \cite{Merzbacher}
\begin{equation}
\label{WF_Coul}
\psi \left(\mathbf{r},t\right) = e^{i p_{\textrm{a}} z -i \eps_{\mathrm{a}} t}~_{1}F_{1} \left[in;1;ik(r-z) \right] ,
\end{equation}
where $n = \frac{G m_{\mathrm{a}} M}{\hbar v_{\mathrm{a}}}$ in SI units and $_{1}F_{1}$ is the confluent hypergeometric function of the first kind. For an axion located at the surface of the Earth, $p_{\textrm{a}} (r-z) \gg 1$ unless $r \approx z$. The large argument expansion of the confluent hypergeometric function in (\ref{WF_Coul}) gives
\begin{align}
\label{WF_Coul_asymp}
\psi \left(\mathbf{r},t\right) &\approx \frac{e^{-n\pi/2} e^{-i \eps_{\mathrm{a}} t}}{\left|\Gamma(1+in)  \right|}
\left\{ e^{i p_{\textrm{a}} z -i n\ln [p_{\textrm{a}}(r-z)] + i\sigma_n} \right. \notag \\ 
&+ \left. \frac{n}{p_{\textrm{a}}(r-z)} e^{i p_{\textrm{a}} r +i n\ln [p_{\textrm{a}}(r-z)] - i\sigma_n} \right\} ,
\end{align}
where the phase factor $\sigma_n$ is defined by $\Gamma(1+in) = \left|\Gamma(1+in) \right| e^{i \sigma_n}$. We again assume that the axion field is classical and, therefore, real. In the non-relativistic limit, $\eps_{\mathrm{a}} \approx m_{\mathrm{a}}$ from the dispersion relation (\ref{ax_disp_reln}). We also define the axion field prefactor to be $a_{0}$ in accordance with Eq.~(\ref{axion_field}). Hence the axion field near the surface of the Earth can be written as
\begin{align}
\label{axion_Coul_field_asymp}
&\phi \left(\mathbf{r},t\right) \approx a_{0}   \left\{ \cos\left(p_{\textrm{a}} z - m_{\mathrm{a}} t - n\ln [p_{\textrm{a}}(r-z)] + \sigma_n\right) \right. \notag \\ 
&+ \left. \frac{n}{p_{\textrm{a}}(r-z)} \cos\left(p_{\textrm{a}} r - m_{\mathrm{a}} t + n\ln [p_{\textrm{a}}(r-z)] - \sigma_n\right) \right\} .
\end{align}
The first term in (\ref{axion_Coul_field_asymp}) is essentially the analogue of the free axion field in Eq.~(\ref{axion_field}), but note the presence of the additional phase factor $n\ln [p_{\textrm{a}}(r-z)]$ in (\ref{axion_Coul_field_asymp}). This additional phase factor remains for a particle in a Coulomb-like potential even in the $r \to \infty$ limit, but the second term in Eq.~(\ref{axion_Coul_field_asymp}) tends to zero in the same limit for a fixed value of $z$. The second term in Eq.~(\ref{axion_Coul_field_asymp}) is responsible for the increase in axion density near the surface of a gravitating body, compared with that at infinitely large distances away from the gravitating body. For an axion located at the surface of the Earth, $\frac{n}{p_{\textrm{a}}r} \approx 10^{-3}$, while for an axion located at the surface of the Sun, $\frac{n}{p_{\textrm{a}}r} \approx 3$. An increased axion density can thus result in an enhancement of axion-induced phenomena.

We now consider a relative of the axion-wind effect of Refs.~\cite{Victor13,Graham13}, which exists only in the presence of a gravitating field. With the aid of the first line of (\ref{H_int_NRL_spat}), the Hamiltonian governing the axion-wind effect in the presence of a gravitational field reads
\begin{widetext}
\begin{align}
\label{axion_wind_grav}
H_{\textrm{int}}^{\textrm{spat}} \left(t\right) = \frac{a_{0} p_{\textrm{a}} \mathbf{\sigma}_{\mathrm{\lambda}}}{f_{\textrm{a}}} \cdot &\left\{\mathbf{\hat{z}}\sin(m_{\textrm{a}} t + \eta') \left[1+ \frac{n}{p_{\textrm{a}}(r-z)} \right] - \mathbf{\hat{r}} \frac{n\sin(m_{\textrm{a}} t + \eta')}{p_{\textrm{a}}(r-z)} + \mathbf{\hat{r}} \frac{n\sin(m_{\textrm{a}} t + \eta'')}{p_{\textrm{a}}(r-z)} \left[1+ \frac{n}{p_{\textrm{a}}(r-z)} \right] \right. \notag \\
& - \left. \mathbf{\hat{z}} \frac{n^2 \sin(m_{\textrm{a}} t + \eta'')}{p_{\textrm{a}}^2 (r-z)^2} + \mathbf{\hat{r}} \frac{n\sin(m_{\textrm{a}} t + \eta'' - \frac{\pi}{2})}{p_{\textrm{a}}^2 (r-z)^2} - \mathbf{\hat{z}} \frac{n\sin(m_{\textrm{a}} t + \eta'' - \frac{\pi}{2})}{p_{\textrm{a}}^2 (r-z)^2}      \right\} ,
\end{align}
\end{widetext}
where the phase factors $\eta'$ and $\eta''$ are in general not equal, and $\mathbf{\sigma}_{\mathrm{\lambda}}$ is the spin operator for an electron ($\lambda=e$) or nucleon ($\lambda=N$) in the atomic, molecular or nuclear system of interest. There are two distinct contributions (direction-wise) to the generalised axion-wind effect in a gravitational field, described by Eq.~(\ref{axion_wind_grav}). The first is directed along the background axionic field's direction of propagation in space and is proportional to $\mathbf{\sigma}_{\mathrm{\lambda}} \cdot \mathbf{p}_{\mathrm{a}}$, while the second is directed along the gravitational field gradient generated by the body of interest and is proportional to $\mathbf{\sigma}_{\mathrm{\lambda}} \cdot \mathbf{g}$. By averaging over the period of the rotation of the Earth about its own axis, the terms proportional to $\mathbf{\sigma}_{\mathrm{\lambda}} \cdot \mathbf{p}_{\mathrm{a}}$ average to zero and only the terms proportional to $\mathbf{\sigma}_{\mathrm{\lambda}} \cdot \mathbf{g}$ remain. After such averaging, one can also search for the $\mathbf{\sigma}_{\mathrm{\lambda}} \cdot \mathbf{g}$ effect. Thus the interaction of the spin of either an electron or nucleon, which also interacts with an axion field, with the gravitational field gradient of a gravitating body can give rise to axion-induced observable effects. The static interaction of the spin of a SM fermion with either a gravitational field gradient or some preferred direction in space has previously been considered, see e.g.~Refs.~\cite{Venema92,Heckel08,Flambaum09SG,Peck12,Romalis13arXiv}.

\section{Conclusions}
\label{Sec:Concl}
We have shown that the interaction of an axion field, or in general a pseudoscalar field, with the axial-vector current generated by an electron through a derivative-type coupling according to the Lagrangian density (\ref{L_int}) can give rise to time-dependent mixing of opposite-parity states in atomic and molecular systems. Likewise, the analogous interaction of an axion field with the axial-vector current generated by a nucleon can give rise to time-dependent mixing of opposite-parity states in nuclear systems. This mixing can induce oscillating EDMs, oscillating PNC effects and oscillating anapole moments in such systems. We suggest that the first two of these effects can be measured by applying a static electric field to the system of interest. We have derived corresponding expressions for such axion-induced EDMs in Group I elements and systems with a single nearly degenerate pair of opposite-parity states. By adjusting the energy separation between the opposite-parity states of interest to match the axion mass energy, axion-induced experimental observables can be enhanced by many orders of magnitude. Measurements of these effects permit either the determination of or the placing of limits on important physical axion parameters, namely $m_{\textrm{a}}$ and $a_{0}/f_{\textrm{a}}$. We have considered the oscillating atomic EDMs that can be generated by axions through hadronic mechanisms, namely the P,T-violating nucleon-nucleon interaction and through the axion-induced EDMs of valence nucleons, the latter of which was considered in Refs.~\cite{Graham11,Graham13}, and derived corresponding expressions for the axion-induced EDM for $^{199}$Hg, which at present provides the most sensitive probe for static EDM measurements in diamagnetic atoms \cite{Romalis09,Romalis13A}, and $^{225}$Ra, which can offer a several order-of-magnitude enhancement in EDM magnitude over that for $^{199}$Hg. Finally, we have shown that the interaction of the spin of either an electron or nucleon, which also interacts with an axion field, with the gravitational field gradient of a gravitating body can give rise to axion-induced observable effects. These effects, which are of the form $\mathbf{g} \cdot \mathbf{\sigma}$, differ from the axion-wind effect, which has the form $\mathbf{p}_{\textrm{a}} \cdot \mathbf{\sigma}$. 

Regarding the choice of system, with which to perform the Stark-inteference experiments described in Sec.~\ref{Sec:pnc}, it is important to bear in mind how the to-be-determined axion mass $m_{\textrm{a}}$ might compare with the energy separation between the opposite-parity levels of interest. Searching for resonance-enhanced axion-induced effects by using a static, external magnetic field to vary the energy separation between the opposite-parity states of interest is one possible strategy. Note that there is no significant advantage in using heavy atomic or molecular (that is, containing at least one heavy atom) systems for such axion detection experiments, unlike in experiments that search for static effects of PNC-mixing of opposite-parity states induced by the neutral weak interaction, where the desired effects scale approximately as $Z^{3}$ \cite{Bouchiat74,Budker08} and so there is an obvious advantage in using heavy atomic or molecular systems for such experiments. Therefore, there are many possible candidate systems for such axion detection schemes. Atomic systems, which may be useful in such axion detection experiments, include Dy, Cs, Yb, Tl, Ra and Ra$^{+}$. Diatomic molecular radical species are particularly advantageous with regard to searches for resonance-enhanced axion-induced effects, since energy separations between opposite-parity states in the range $10^{-6} - 1$ eV, in which the axion mass is currently believed to lie, are easily achieved in such species. Compared with atomic species, the levels of diatomic molecular radical species also have quite narrow natural widths, which should enhance the axion-induced resonance signal to a greater degree in molecular systems. A further advantage of molecular radical species, such as SrF, ZrN, BaF, YbF, AlS, GaO, MgBr, LaO, PbF and ThO, is that they are already considered for high precision experiments to study violations of the fundamental symmetries of nature (see e.g.~Refs.~\cite{DeMille08,Alphei11,ThO13}). Finally, we mention that a significant reduction in relative statistical error may be achieved in solid-state experiments. Static electron EDM experiments in ferroelectrics are discussed in Refs.~\cite{Eckel12,Budker06}, for instance. Several detection schemes have also recently been proposed to detect axionic dark matter in solid-state systems \cite{CASPEr2014,Beck2013}.

\section*{ACKNOWLEDGEMENTS}
We are grateful to M.~Yu.~Kuchiev, D.~Budker, M.~G.~Kozlov and D.~DeMille for important discussions. This work is supported by the Australian Research Council.



\end{document}